\documentclass[useAMS,usenatbib]{mn2e}
\usepackage{epsf}

\def\Msun{{\rm M_\odot}}

\title[The survival and disruption of CDM micro-haloes]
{The survival and disruption of CDM micro-haloes: implications for 
direct and indirect detection experiments}

\author[Tobias Goerdt et al.]
{\parbox[t]{\textwidth}{Tobias Goerdt$^1$\thanks{tgoerdt@physik.unizh.ch}, 
 Oleg Y. Gnedin$^2$,
 Ben Moore$^1$,
 J\"urg Diemand$^3$ \&\\
 Joachim Stadel$^1$}\\  \vspace*{3pt} \\
$1$ Institut f\"ur Theoretische Physik, Universit\"at Z\"urich,
    Winterthurerstrasse 190, CH-8057 Z\"urich, Schweiz\\
$2$ The Ohio State University, Department of Astronomy,
    140 W 18th Avenue, Columbus, OH 43210, USA\\
$3$ University of California, Department of Astronomy and Astrophysics,
    1156 High Street, Santa Cruz CA 95064, USA}

\date{Draft version \today}
\pagerange{\pageref{firstpage}--\pageref{lastpage}}
\pubyear{2006}

\begin{document}

\maketitle

\label{firstpage}

\begin{abstract}
If the dark matter particle is a neutralino then the first structures to form
are cuspy cold dark matter (CDM) haloes collapsing after redshifts $z\approx
100$ in the mass range $10^{-6}-10^{-3}\, \Msun$. We carry out a detailed
study of the survival of these micro-haloes in the Galaxy as they experience
tidal encounters with stars, molecular clouds, and other dark matter
substructures.  We test the validity of analytic impulsive heating calculations
using high resolution $N$-body simulations. A major limitation of analytic
estimates is that mean energy inputs are compared to mean binding energies,
instead of the actual mass lost from the system.  This energy criterion leads
to an overestimate of the stripped mass and underestimate of the disruption
timescale since CDM haloes are strongly bound in their inner parts. We show
that a significant fraction of material from CDM micro-haloes can be unbound by
encounters with Galactic substructure and stars, however the cuspy central
regions remain relatively intact. Furthermore, the micro-haloes near the solar
radius are those which collapse significantly earlier than average and will
suffer very little mass loss. Thus we expect a fraction of surviving bound
micro-haloes, a smooth component with narrow features in phase space, which may
be uncovered by direct detection experiments, as well as numerous surviving
cuspy cores with proper motions of arc-minutes per year, which can be detected
indirectly via their annihilation into gamma-rays.
\end{abstract}

\begin{keywords}
cosmology: theory -- dark matter -- galaxies: formation -- 
gamma-rays: theory -- methods: numerical 
\end{keywords}

\section{Introduction}

If dark matter is composed mainly of the lightest supersymmetric partner
particle, the neutralino, the first self-gravitating structures in the Universe
are Earth-mass haloes forming at high redshifts
\citep*{hofmann_etal01,diemand05}. As many as $10^{15}$ could be within our
Galactic halo today. These abundant cold dark matter (CDM) micro-haloes have
cuspy density profiles that can withstand the Galactic tidal field at the solar
radius. The numbers of such haloes that lie within the vicinity of the solar
system depends on how many survive the complex merging history of early
hierarchical structure formation. $N$-body simulations of CDM satellites
indicate that tightly bound cusps are very stable against tidal stripping
(Kazantzidis et al. 2004), and therefore dense micro-haloes accreting late onto
more massive structures may survive relatively intact. The exact distribution
of dark matter in the solar vicinity is important for direct and indirect dark
matter detection experiments.

Substructures that survive the merging process will experience continuous
perturbative encounters with stars, molecular clouds, and other dark matter
subhaloes. As discussed in \citet{diemand05}, we expect that these encounters
lead to some mass loss but that the cusps of most micro-haloes remain intact.
Recent studies by \citet{zhao_etal05a, zhao_etal05b}, \citet{green_goodwin06} 
and \citet{berezinsky} have raised the question whether these first haloes
would be completely disrupted by close encounters with stars. Crossing the
Galactic disc would also cause additional tidal heating. \citet{moore_etal05}
argued that the analytical impulse approximation and the semi-analytic models
used in these studies may not fully describe the disruption of the high-density
inner cores. Particle orbits deep in the cusp may remain adiabatically
invariant to the perturbations and preserve the structure of the cusp. Only
direct numerical simulations can describe these complex dynamical processes. In
this paper, we use several sets of high resolution $N$-body simulations to
test the validity of analytical heating models.

An important factor in the survival statistics of micro-haloes is how many
survive similar-mass mergers during the build-up of the Galactic halo
\citep{diemand2006}. Even if only a few percent survive the hierarchical
growth, many micro-haloes would still lie within one parsec from the Sun. Their
dense cuspy cores would be sources of gamma-ray emission due to
self-annihilation, which could be uniquely distinguished by their high proper
motions on the sky of the order arc-minutes per year (Moore et al. 2005,
Koushiappas 2006).

\section{Heating by stars in the solar neighbourhood}
  \label{sec:imp}

There are various ways to define a virialized halo. The approach often used in
cosmological simulations, which we adopt here, is that dark haloes virialize
when their average density equals $\Delta = 200$ times the mean density of the
Universe, $\bar{\rho}(z) = 3 \Omega_0 H_0^2 (1+z)^3 / 8\pi G$.  Here $H_0$ is
the Hubble constant, $\Omega_0$ is the matter density parameter, and $z$ is the
redshift of virialization. Then the virial radius of the halo, defined by the
relation $M_{\rm vir} \equiv {4\pi \over3} R_{\rm vir}^3 \Delta \bar{\rho}(z)$,
is
\begin{equation}
  R_{\rm vir} = 0.31 \, (1+z)^{-1} 
     \left({M_{\rm vir} \over 10^{-6}\, \Msun}\right)^{1/3}
     \mbox{pc},
\end{equation}
for $\Omega_0 = 0.3$ and $H_0 = 70 \, \mbox{km s}^{-1} \, \mbox{Mpc}^{-1}$. The
virial velocity is defined by the relation $V_{\rm vir}^2 \equiv G M_{\rm vir}
/R_{\rm vir}$:
\begin{equation}
  V_{\rm vir} = 12 \, (1+z)^{1/2} 
       \left({M_{\rm vir} \over 10^{-6}\, \Msun}\right)^{1/3} 
       \mbox{cm s}^{-1}.
  \label{eq:vvir}
\end{equation}
These parameters determine the binding energy of the haloes, which can be
expressed using the half-mass radius of the system: $E_{\rm b} \approx 0.2 G
M_{\rm vir}/R_{\rm 1/2}$ \citep{spitzer87}. Density profiles of dark matter
haloes in cosmological simulations are often described by the NFW model with a
concentration parameter, $c$.  For $c < 10$, the radius containing half of the
virial mass is approximately $R_{1/2} \approx (5c)^{-1/4} \, R_{\rm vir}$.
High-redshift haloes have typically low concentrations, such that $R_{1/2}
\approx 0.5 R_{\rm vir}$.  Therefore, the binding energy of first haloes is
$E_{\rm b} \approx 0.4 V_{\rm vir}^2$.

As these small haloes merge into larger systems, two effects may modify their
structure: tidal truncation by the host galaxy and tidal heating by massive,
fast-moving perturbers (stars, molecular clouds, other dark matter
substructures).

In the vicinity of the Sun, the matter density is dominated by stars, which we
assume to have the same mass $m_* = 0.7\, \Msun$.  The stellar mass density is
$m_* n_* \approx 0.1 \, \Msun$\,pc$^{-3}$ \citep{BM98}, which is a half of the
total density of the disc calculated from the Oort limit \citep{bahcall84}. In
order to remain gravitationally self-bound, the micro-haloes must have an
average density above roughly $2 m_* n_*$.

Fast encounters with massive perturbers increase the velocity dispersion of
dark matter particles and reduce a halo's binding energy.  A distant encounter
at an impact parameter $b$ with a relative velocity $V_{\rm rel}$ increases the
energy per unit mass on the average by
\begin{equation}
  \Delta E_1(b) \approx {1\over 2}
    \left({2 G m_* \over b^2 \, V_{\rm rel}}\right)^2
    {2\over 3} \left< r^2 \right>,
  \label{eq:de1}
\end{equation}
where $\left< r^2 \right> \sim R_{1/2}^2$ is the ensemble average of the
particle distance squared from the centre of the micro-halo.

At very small impact parameters, $b < b_1$, a single encounter would be
sufficiently strong to unbind the whole halo: $\Delta E_1(b_1) = E_{\rm b}$. As
we show later in section \ref{sec:num}, a small central part always survives
even such a strong perturbation, apart from direct collisions with $b = 0$.
Nevertheless, it is instructive to define the disruptive encounter threshold,
which is given by
\begin{equation}
 b_1 = a_{\rm c}
     \left({G m_* R_{\rm vir} \over V_{\rm rel} V_{\rm vir}}\right)^{1/2}
     \approx 0.2 \, (1+z)^{-3/4} \ {\rm pc},
\end{equation}
where $a_{\rm c} \approx 0.96 \, (c/3)^{-1/8}$. Equation (\ref{eq:de1}) is
strictly valid only in the tidal approximation, $b \gg R_{\rm vir}$. An
encounter at $b_1$ falls in that regime for redshifts $z < 50$, which is
appropriate for our consideration of the micro-haloes.

The number of encounters over time $t$ as a function of impact parameter is
$dN_{\rm enc}(b) = n_* V_{\rm rel} t \, 2\pi b db$, where $n_*$ is the number
density of stars.  We can obtain the cumulative effect of multiple
non-disruptive encounters by integrating over the impact parameter:
\begin{eqnarray}
  \Delta E_{\rm tid} 
     & = & \int_{b_1}^{b_{\rm max}} 
           \Delta E_1(b) \, {dN_{\rm enc} \over db}\, db \nonumber\\
     & = & 0.4 a_{\rm c}^4 \pi
           {G^2 m_*^2 R_{\rm vir}^2 n_* t \over V_{\rm rel}} 
           \left({1 \over b_1^2} - {1 \over b_{\rm max}^2}\right).
  \label{eq:etid}
\end{eqnarray}
The upper limit of integration is set by the condition that the encounter is
impulsive, i.e. the duration of the encounter $b/V_{\rm rel}$ is shorter than
the orbital time of particles in the halo, $R_{\rm vir}/V_{\rm vir}$. The
maximum impact parameter is given by
\begin{equation}
  \left({b_{\rm max} \over b_1}\right)^2 \approx
   a_{\rm c}^2{V_{\rm rel}^3 \, R_{\rm vir} \over G m_* V_{\rm vir}} \gg 1.
\end{equation}
The ratio of the tidal heating energy in non-disruptive encounters to the
binding energy is
\begin{equation}
  {\Delta E_{\rm tid} \over E_{\rm b}} = 
     a_{\rm c}^2 \pi {G m_* n_* t R_{\rm vir} \over V_{\rm vir}}.
  \label{eq:etideb}
\end{equation}

We can also calculate the effect of disruptive encounters, with $b < b_1$. The
number of such encounters is simply
\begin{equation}
  N_{\rm enc}(<b_1) = \pi b_1^2 n_* V_{\rm rel} t
    = a_{\rm c}^2 \pi {G m_* n_* t R_{\rm vir} \over V_{\rm vir}}.
\end{equation}
This number is the same as equation (\ref{eq:etideb}) meaning that the
cumulative effect of disruptive encounters is the same as that of
non-disruptive encounters. The total disruption probability, $N_{\rm tot}$, is
then twice that given by equation (\ref{eq:etideb}).

To calculate this disruption probability, we note that while stars in the solar
neighbourhood move on approximately circular orbits around the Galactic centre,
small dark matter haloes would be moving on isotropic orbits inclined with
respect to the Galactic disc.  Their expected vertical velocity is $V_z \approx
200$ km s$^{-1}$. The crossing time of the disc with a scale height of $H =
0.2$ kpc is $2H/V_z = 2\times 10^6$ yr. In the solar neighbourhood, haloes
would cross the disc every $10^8$ yr and have about 100 crossings in the Hubble
time. The total amount of time the haloes would spend in the region of high
stellar density $m_* n_*$ is then $t_d \sim 2\times 10^8$ yr.  The total
disruption probability is
\begin{eqnarray}
  N_{\rm tot} & = & 2 \, N_{\rm enc}(<b_1) \nonumber\\
    & = & \left({1+z \over 131}\right)^{-3/2} 
      \left({m_* n_* \over 0.1 \, \Msun \, {\rm pc}^{-3}}\right)
      \left({t_d \over 2\times 10^8 \, {\rm yr}}\right).
\end{eqnarray}
Therefore, the haloes virialized after redshift $z = 130$ should suffer 
significant mass loss by passing stars in the solar neighbourhood. Due to
biased halo formation typical subhaloes in the solar neighbourhood come form $2
\sigma$ fluctuations (Diemand, Madau \& Moore 2005), i.e. they virialize at
half the expansion factor (or twice the (z + 1) value) than typical haloes of
the same mass in the field (i.e. $1 \sigma$ peaks). A formation time of z = 130
corresponds to about a $3 \sigma$ peak. Such early structure formation is not
uncommon in dense environments, for example the small, over dense region
simulated in Diemand, Kuhlen \& Madau (2006) already contains 845 micro-haloes
at z = 130. A fraction of about 20\% of the local mass comes from peaks above
$3\sigma$ (Diemand, Madau \& Moore 2005), implying that approximately 20\% of
the local subhalo population should therefore not suffer significant mass loss.

\section{Heating by dark matter substructure}

Virialized, self-gravitating subhaloes within larger haloes (the substructure)
will also kinematically heat and disrupt their small cousins [c.f. Boily et al.
(2004)]. $N$-body simulations \citep{diemand04} show that the number of
subhaloes with masses above $M$ within a host of mass $M_{\rm host}$ scales as
\begin{equation}
  N_{\rm sub}(>M/M_{\rm host}) \approx 
    \left({M \over 10^{-2} \, M_{\rm host}}\right)^{-1}.
  \label{nsub}
\end{equation}
Since stars occupy only a small fraction of the volume of their host haloes, it
is important to consider if the tidal heating by dark matter subhaloes can
disrupt a significant fraction of micro-haloes.

The analysis of section \ref{sec:imp} can be generalised for perturbers with a
range of masses, $M_{\rm vir} < M < 10^{-2}\, M_{\rm host}$.  Let $f \equiv
M/M_{\rm host}$ be the dimensionless subhalo mass. The threshold impact
parameter, at which a single encounter with subhalo $f$ is disruptive, is
$b_1^2(f) \approx f V_{\rm host} R_{\rm host} R_{\rm vir}/V_{\rm vir}$, where
we assumed the relative velocity to be the virial velocity of the host halo,
$V_{\rm rel} \approx V_{\rm host}$.  However, for most subhaloes this impact
parameter is smaller than their size, $R_{\rm sub} \approx r [M_{\rm sub}/
3M_{\rm host}(r)]^{1/3}$, which is determined by tidal truncation at distance
$r$ from the centre of the host halo.  Tidal approximation applies only at $b >
b_{\rm min} = R_{\rm sub}$. Therefore, most encounters will be non-disruptive.

The cumulative heating by multiple non-disruptive encounters with subhaloes of
mass $M_{\rm sub} = f M_{\rm host}$ is [see equation (\ref{eq:etid})]:
\begin{equation}
  {\Delta E_{\rm tid}(f) \over E_{\rm b}} = {\pi G^2 M_{\rm sub}^2 R_{\rm
      vir}^2 t \over V_{\rm host} V_{\rm vir}^2 b_{\rm min}^2} {dn_{\rm sub}
    \over df},
\end{equation}
where $dn_{\rm sub} \over df$ is the number density of subhaloes $f$. Taking an
NFW model for the smooth component of the Galactic halo and restricting our
analysis to the inner part of the halo, $r \la r_s \approx 20$ kpc, we find the
subhalo's truncation radius $R_{\rm sub} \approx r_s (f r / r_s)^{1/3}$.  The
density of subhaloes, assuming they have not been completely disrupted, is
\begin{equation}
  {dn_{\rm sub} \over df} \approx {10^{-2} f^{-2} \over 4\pi g(c) r_{\rm s}^2
    r},
\end{equation}
where $g(c) \equiv \ln{(1+c)} - c/(1+c) \approx 1.6$ for a concentration
parameter $c = 12$ \citep{klypin_etal02}. For the Galaxy, a Hubble time
corresponds to $t \sim 5 R_{\rm host}/V_{\rm host}$.

Thus for the inner halo,
\begin{eqnarray}
  {\Delta E_{\rm tid}(f) \over E_{\rm b}} & = & {5\times 10^{-2} c^5 \over 4
    g(c)} \left({t V_{\rm host} \over 5 R_{\rm host}}\right) 
   {R_{\rm vir}^2 V_{\rm host}^2 \over R_{\rm host}^2 V_{\rm vir}^2} \nonumber
   \\ & \times & f^{-2/3} \left({r \over r_{\rm s}}\right)^{-5/3}.
\end{eqnarray}

Integrating over all subhaloes, $f < 0.01$, we find
\begin{equation}
  {\Delta E_{\rm tid} \over E_{\rm b}} \approx 0.063
   \left({r \over r_{\rm s}}\right)^{-5/3}.
\end{equation}
Thus, mini-haloes may be disrupted by repeated encounters with more
massive haloes within $r \la 0.2 r_{\rm s} \approx 4$ kpc from the centre of
the Galaxy.

\section{Numerical tests of the impulsive approximation}
  \label{sec:num}

In this section we test the response of a CDM micro-halo to repeated impulsive
encounters using N-body calculations in order to test the validity of the
impulse approximation, and to study in detail how the internal structure of the
micro-haloes evolve with time.

The initial state for the micro-halo is an equilibrium profile with the same
structural parameters as found by \citet{diemand05} at $z = 26$, the epoch at
which such structures are typically accreted into larger mass systems. This
halo obeys a cuspy density profile, the general $\alpha, \beta, \gamma$ law
\citep{abc}:
\begin{equation}
\rho(r)=\frac{\rho_0} {\left({\frac{r}{r_{\rm s}}}\right)^\gamma \left[{1 +
\left({\frac{r}{r_{\rm s}}}\right)^{\alpha}}\right]^{\frac{\beta - \gamma}
{\alpha}}} \qquad (r \le R_{\rm micro}),
\label{eq:den}
\end{equation}
with $\alpha = 1.0$, $\beta = 3.0$ and $\gamma = 1.2$. The mass of the halo is
$M_{\rm micro} = 10^{-6} \, \Msun$ within the z = 26 virial radius $R_{\rm
micro} = 0.01$ pc. The concentration parameter is low, $R_{\rm micro}/r_{\rm s}
= 1.6$, typical of micro-haloes in the field at z = 26. Some experiments we
repeated with micro-haloes having concentrations of 3.2. The typical local
subhalo forms earlier (by about a factor of two in redshift, see Diemand, Madau
\& Moore 2005) than the average micro-halo in the field. Therefore the typical
local subhalo might be twice as concentrated and more robust against mass loss.
However, to be conservative we use the low concentration of 1.6 throughout this
paper unless stated otherwise. For numerical stability of the profile, we make
a realization of this halo extending to approximately $4 \, R_{\rm micro}$
using the techniques of \citet{kazantzidis_etal04}.  At $r > R_{\rm micro}$,
the density profile falls off exponentially as $\exp(-r/r_{\rm decay})$, with
$r_{\rm decay} = 0.3 \, R_{\rm micro}$. The total mass of the system is
therefore $1.81 \, M_{\rm micro}$. We use $10^6$ particles of equal mass,
$m_{\rm p} = 1.81 \times 10^{-12} \Msun$.  The force calculations have a
softening length of $0.005 \, R_{\rm micro}$.

We then subject the equilibrium micro-halo to a series of impulsive encounters
with a star of mass $m_* = 0.7 \, \Msun$, the mean mass per star in the disc of
the Galaxy. First we run six simulations, which differ in the minimal distance
from the star to the centre of the micro-halo.  For these six simulations the
impact parameters are $b = 0.005, 0.01, 0.02, 0.05, 0.1, 1$\,pc. In all runs
the star moves with the relative velocity $V_{\rm rel} = 300$ km s$^{-1}$. The
initial separation of the star and the halo along the direction of motion is
three times the impact parameter or three times the micro-halo radius of the
halo, whichever is the greater.  After the star reaches the point of closest
approach, we let it move away the same distance from the halo. Then we remove
the star and let the system evolve in isolation for $3\times 10^8$ yr, which
corresponds to 20 crossing times of the halo. Similar experiments date back to
e. g. \citet{aguilar85}.

Each encounter increases the internal energy of the micro-halo. Following the
perturbation, the system undergoes a series of virial oscillations (contraction
and expansion) until the potential relaxes into a new equilibrium configuration
\citep{gnedin_ostriker99}. Depending on the strength of the perturbation, this
potential relaxation takes between 10 and 20 crossing times of the halo. The
final configuration has a lower binding energy and some particles may escape
the system entirely.

\begin{figure}
\begin{center}
\epsfysize=8.4cm
\epsffile{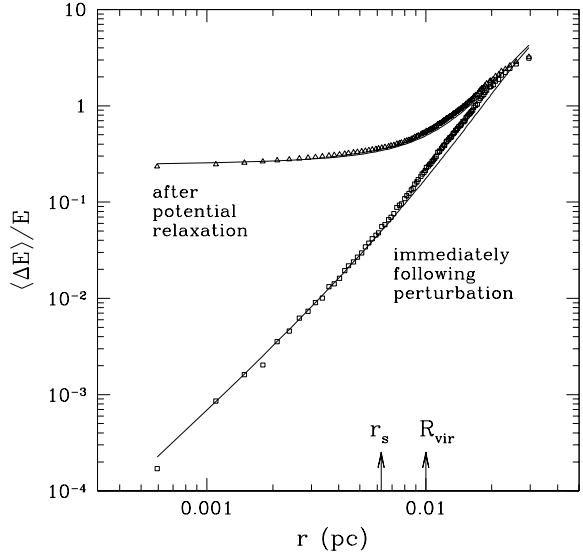}
\end{center}
\caption{Energy change per particle as a function of radius
  immediately following the perturbation ({\it squares}) and after
  potential relaxation ({\it triangles}), for the encounter with $b =
  0.02$ pc.}
  \label{fig:f1}
\end{figure}

Fig. \ref{fig:f1} shows the energy change per particle for a flyby at $b =
0.02$ pc at two different times: directly following the encounter and after the
potential relaxation.  The duration of the encounter, $\tau = 2b/V_{\rm vel}
\approx 130$ yr, is much shorter than the dynamical (crossing) time of the
particles in the micro-halo, $t_{\rm dyn} \approx 1.5\times 10^7$ yr.
Therefore, we expect that the tidal heating can be calculated in the impulsive
approximation [c.f. equation (\ref{eq:de1})]. For ensemble-average of stars
with initial energy $E$, the energy per unit mass increases by the amount
\begin{equation}
  \left<\Delta E\right> = {4\over 3}
    \left({G m_* \over b^2 \, V_{\rm rel}}\right)^2 r^2.
  \label{eq:de_av}
\end{equation}
This prediction is plotted next to the numerical result (squares) in Fig.
\ref{fig:f1} and agrees with it very well.

Subsequent potential relaxation reduces the depth of the potential well of the
system, leading to another effective energy change. \citet{gnedin_ostriker99}
found that this additional energy change can be approximated as a constant
fraction of the initial potential, $\Phi_{\rm i}$:
\begin{equation}
  \Delta E_{\rm pot}(r) = c \, (-\Phi_{\rm i}(r)),
  \label{eq:de_pot}
\end{equation}
where the constant $c$ is such that the sum of $\Delta E_{\rm pot}(r)$ over all
particles is twice the initial energy change of the system, $\Delta E_1(b)$, as
required by the virial theorem. The final energy difference is $\left<\Delta
E\right> + \Delta E_{\rm pot}$. This prediction is plotted next to the
numerical result (triangles) in Fig. \ref{fig:f1}
and again provides a very good fit.

\begin{figure}
\begin{center}
\epsfysize=8.4cm
\epsffile{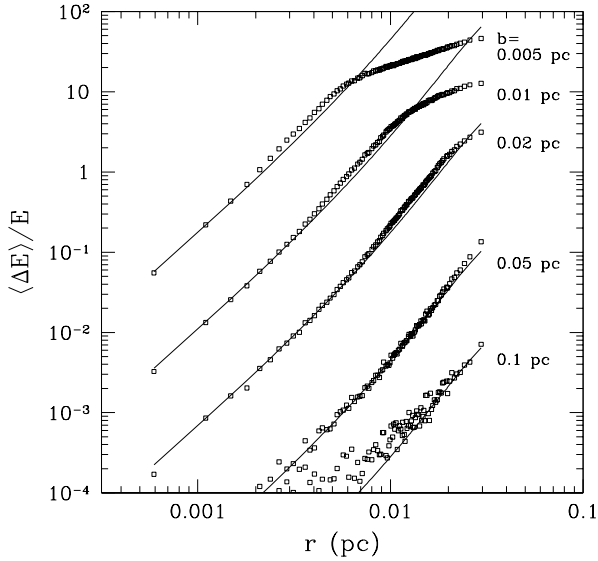}
\end{center}
\caption{Energy change per particle as a function of radius immediately
  following the perturbation, for encounters with different impact
  parameters $b$.}
  \label{fig:f2}
\end{figure}

Fig. \ref{fig:f2}
shows the energy changes immediately following the encounters at different
impact parameters, $b$.  The analytical formula (\ref{eq:de_av}) provides a
good description of the numerical results, except in cases of extremely strong
perturbations when the energy changes by more than 100\%, $\left<\Delta
E\right> > |E|$.

\begin{figure}
\begin{center}
\epsfysize=8.4cm
\epsffile{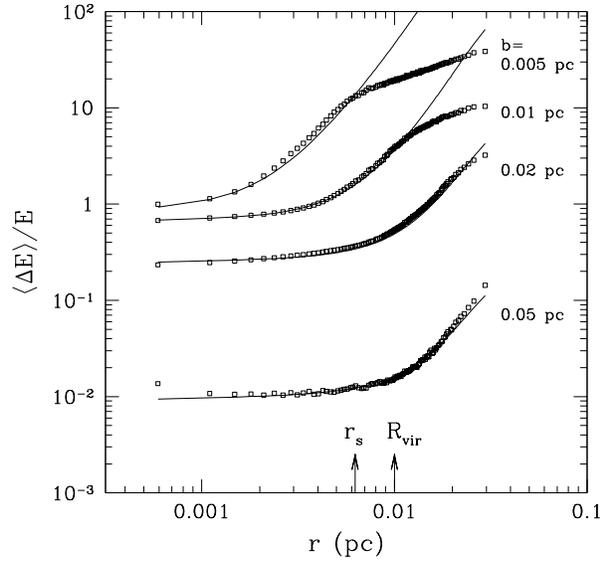}
\end{center}
\caption{Energy change per particle as a function of radius after
  potential relaxation, for encounters with different impact
  parameters $b$.}
  \label{fig:f3}
\end{figure}

Fig. \ref{fig:f3}
shows the final energy redistribution after potential relaxation, for
encounters with different impact parameters, $b$.  Equations (\ref{eq:de_av})
and (\ref{eq:de_pot}) describe the effect very accurately.

\begin{figure}
\begin{center}
\epsfxsize=8.4cm 
\epsffile {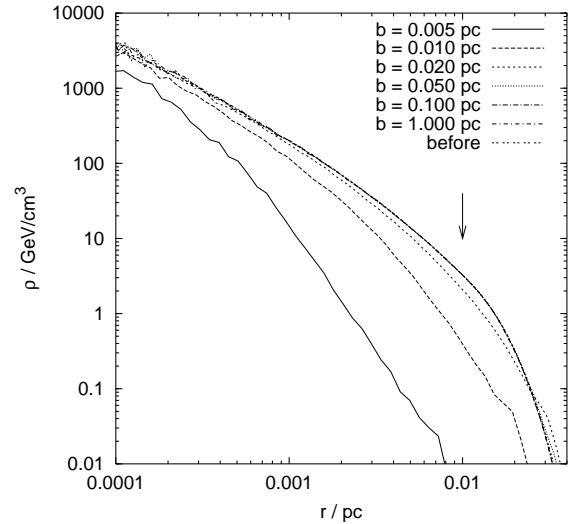}
\end{center}
\caption{Density profile of the micro-halo in a new equilibrium, after
  encounters with different impact parameters, $b$. The arrow indicates the
micro-halo radius.}
  \label{fig:den}
\end{figure}

The change of the micro-halo potential leads to the change of the density
profile.  Particles that gain enough energy escape the system form unbound
tidal tails.  The final density profiles for the encounters with different
impact parameters are shown in Fig. \ref{fig:den}.

The amount of mass stripped from the halo depends on the definition of the
bound mass.  The density of the outer halo profile extends as $r^{-3}$ beyond
the nominal micro-halo radius, and all of the particles are initially bound. We
use two practical definitions. (i) We have defined an effective maximum radius,
$R_{\rm max} \equiv 4 R_{\rm micro}$, beyond which all particles are assumed to
be lost of the micro-halo.  In practice, this radius can be set by the external
tidal field. (ii) We have also defined an effective tidal potential at that
radius, $\Phi_{\rm t} \equiv \Phi(R_{\rm max})$.  We use this tidal potential
to construct another definition of unbound particles, as those with $E >
\Phi_{\rm t}$.  After the potential relaxation, the new potential $\Phi$ is
used to define $\Phi_{\rm t}$ at the same fixed radius $R_{\rm max}$.

\begin{figure}
\begin{center}
\epsfxsize=8.4cm
\epsffile{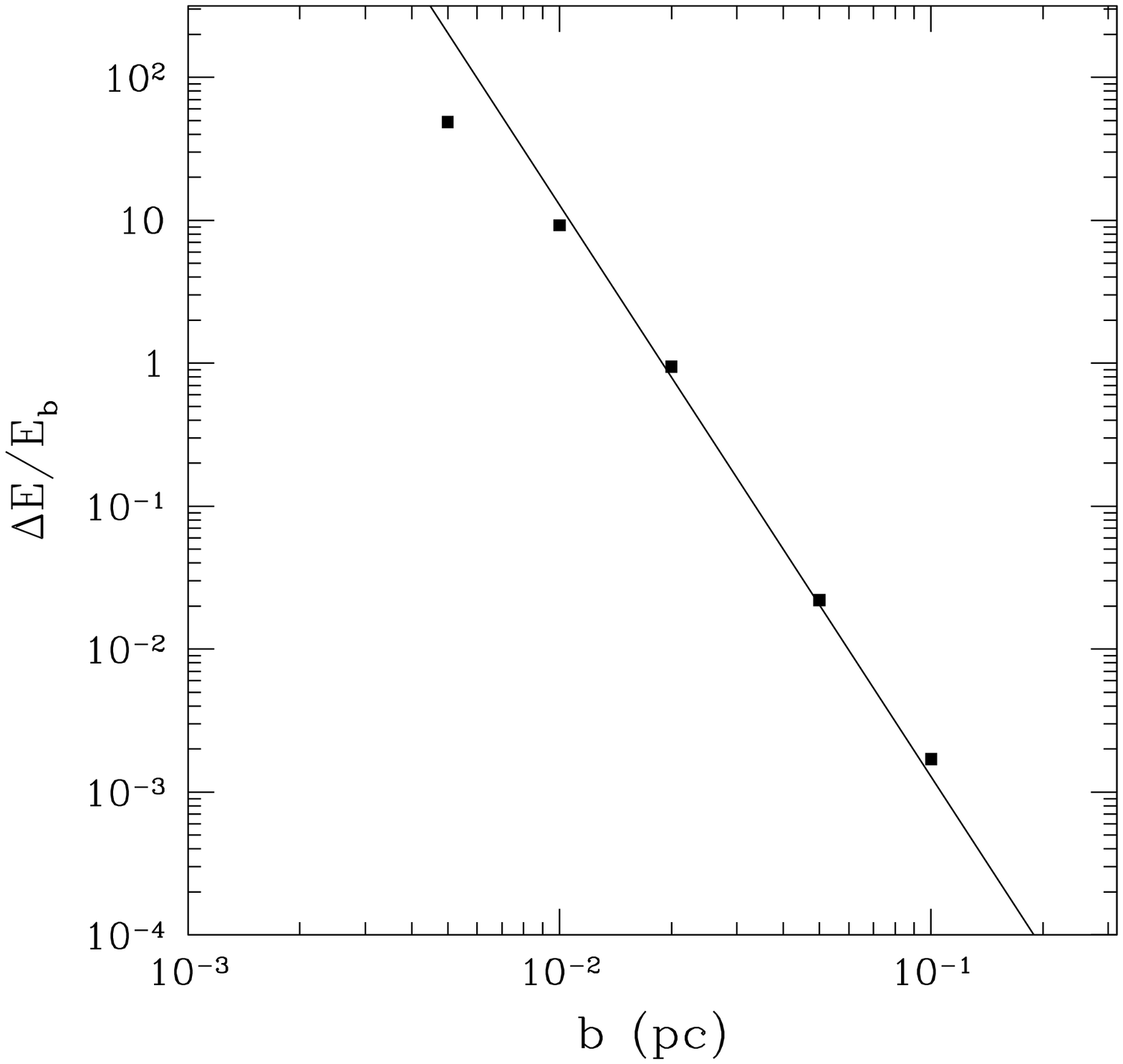}
\end{center}
\caption{Total energy change of all particles immediately following
  the perturbation, as a function of impact parameter ({\it filled
  squares}) and the analytical prediction in the impulse approximation
  ({\it solid line}).}
  \label{fig:de_tot}
\end{figure}

Fig. \ref{fig:de_tot}
shows the change of the total energy of the system immediately following the
encounter, for all particles within $R_{\rm max}$.  It is well described by
equation (\ref{eq:de1}), which is plotted as a solid line.  Since the density
profile of the system continues beyond the nominal micro-halo radius, the
average radius for all particles is $\left< r^2 \right> = 1.63 \,
R_{\rm micro}^2$. We used this value for the plotted prediction.

\begin{figure}
\begin{center}
\epsfxsize=8.4cm
\epsffile{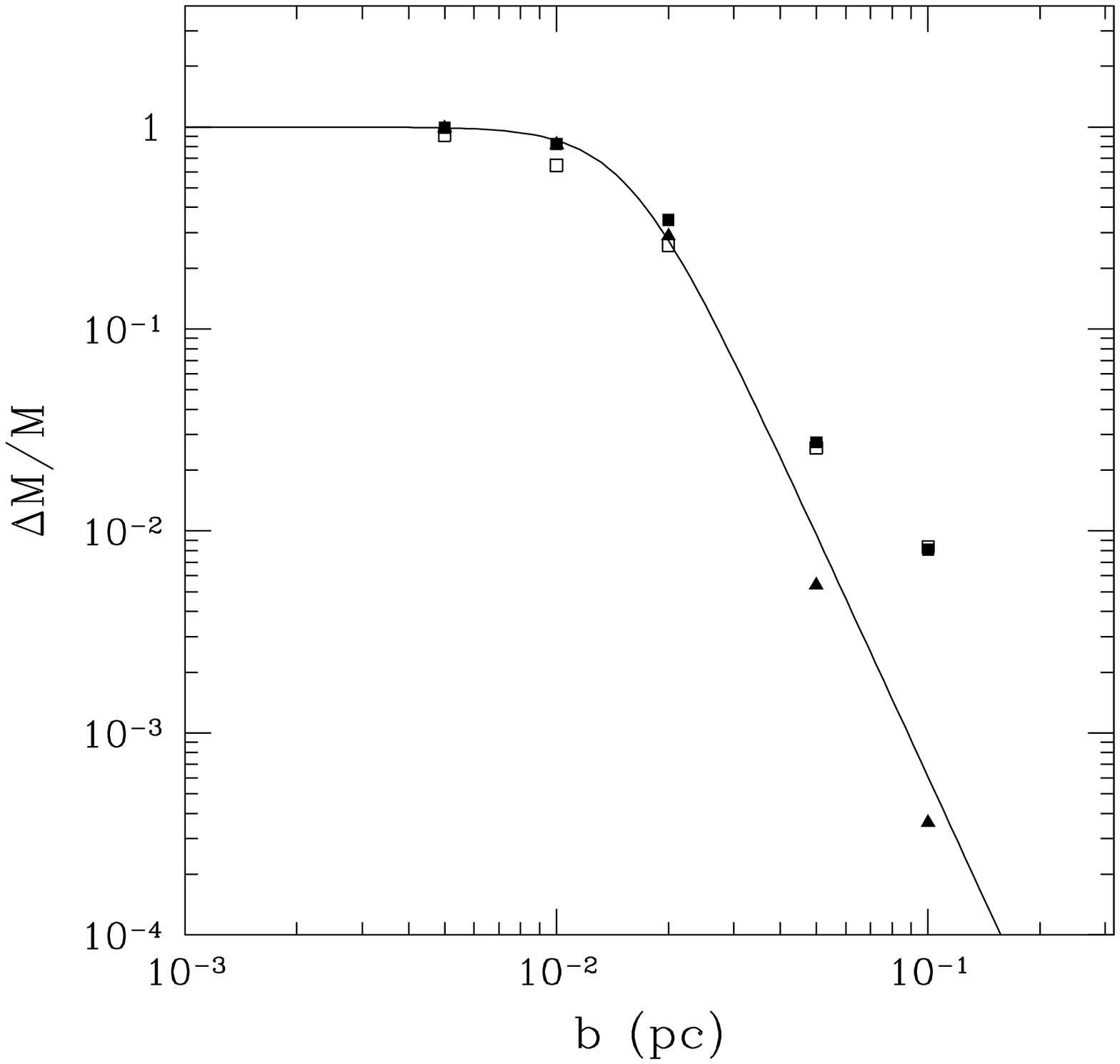}
\end{center}
\caption{Mass loss of the halo as a function of the impact parameter
  $b$, determined using either the energy criterion, $E > \Phi_{\rm
  t}$ ({\it open squares}: immediately following the perturbation,
  {\it filled squares}: after the potential relaxation) or the
  position criterion, $r > R_{\rm max}$, after the potential
  relaxation ({\it triangles}).}
  \label{fig:dm}
\end{figure}

Fig. \ref{fig:dm}
shows the mass loss as a function of impact parameter, using the two
definitions based on the position and energy criterion, respectively. For large
impact parameters (weak perturbations), the position criterion indicates
systematically lower mass loss than the energy criterion. Therefore, some
particles within $R_{\rm max}$ may be unbound at the end of the simulation. In
the strong perturbation regime, both criteria give similar results.

While the total energy change of the system can be computed with sufficient
accuracy using the impulsive approximation, the amount of mass lost cannot.
Using our numerical simulations, we seek to establish a practical relation
between $\Delta M/M$ and $\Delta E/E_{\rm b}$.  We find that the following
equation provides a good fit to the numerical results:
\begin{equation}
  \frac{\Delta M}{M} = \frac{1}{1 + 2.1 \left(\Delta E \over E_{\rm b}
    \right)^{-1}}.
  \label{eq:dm}
\end{equation}
For weak perturbations the mass loss scales as the energy change, $\Delta M/M
\approx 0.5 \Delta E/E_{\rm b}$. For very strong encounters, the mass loss
asymptotically approaches unity. Note however, that even for very small impact
parameters, when $\Delta E/E_{\rm b} \gg 1$, a small fraction of the mass
always remains bound, $\approx 2 (\Delta E/E_{\rm b})^{-1}$.

A note on notation. Strictly speaking, the tidal approximation which was used
to derive equation (\ref{eq:de1}) is only valid for $b \gg R_{\rm micro}$. At
smaller impact parameters the energy change does not scale as $b^{-4}$, but it
can be calculated in the opposite asymptotic limit \citep[e.g,][]{moore93,
carr_sakellariadou99, green_goodwin06}. We take an alternative approach by
parametrising the mass loss in our numerical simulations (equation
[\ref{eq:dm}]) based on the formal extrapolation of equation (\ref{eq:de1}) to
all values of $b$.

\begin{figure}
\begin{center}
\epsfxsize=8.4cm
\epsffile{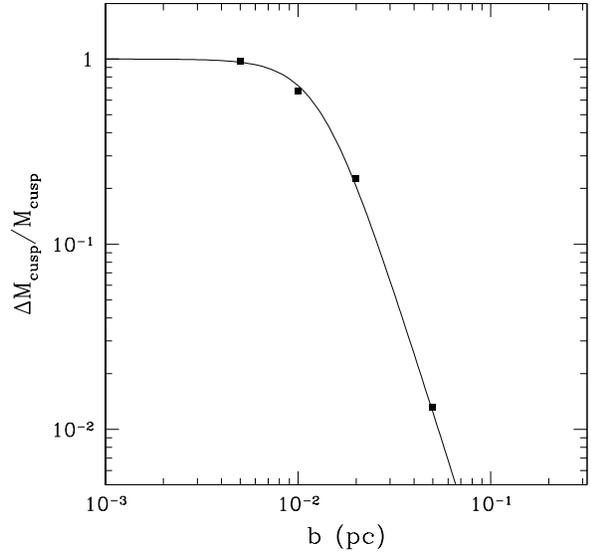}
\end{center}
\caption{Mass lost inside the scale radius, $r < r_{\rm s}$, after the
  potential relaxation, as a function of impact parameter.}
  \label{fig:dm_cusp}
\end{figure}

As Fig. \ref{fig:den} shows, most of the mass remaining bound to the micro-halo
after strong perturbations is concentrated near its centre. It is therefore
interesting to calculate the fraction of lost mass that was initially contained
within the power law density cusp, at $r < r_{\rm s}$.  Fig. \ref{fig:dm_cusp}
shows that this fraction scales as $\Delta M_{\rm cusp}/M_{\rm cusp} \approx (1
+ 5 (\Delta E/E_{\rm b})^{-1})^{-1}$.

We also performed another set of simulations, by repeatedly perturbing the
micro-halo with the same star, with the same relative velocity and at the same
impact parameter $b = 0.02$ pc.  After the halo has relaxed following the first
encounter, we move the centre of mass of the remaining halo to the origin of
the coordinate system and let a star pass by in exactly the same way and again
let it relax, and then repeat the encounter a total of ten times.

\begin{figure}
\begin{center}
\epsfxsize=8.4cm
\epsffile{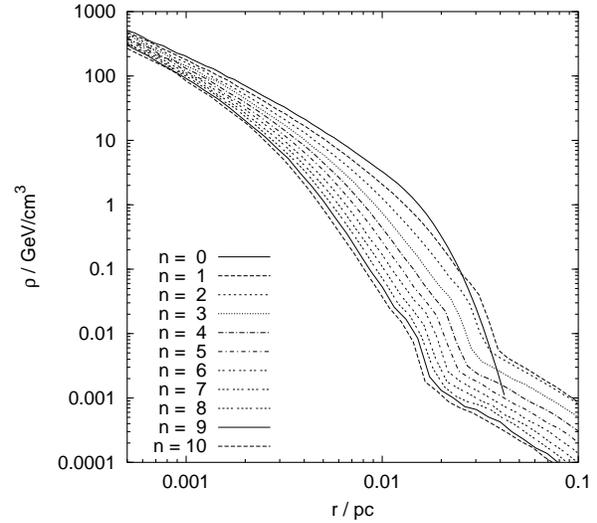}
\end{center}
\caption{Density profile of the micro-halo after successive encounters
  with a star with the same mass and orbital parameters ($b = 0.02$
  pc), marked by the encounter number $n$.}
  \label{fig:den_mult}
\end{figure}

\begin{figure}
\begin{center}
\epsfxsize=8.4cm
\epsffile{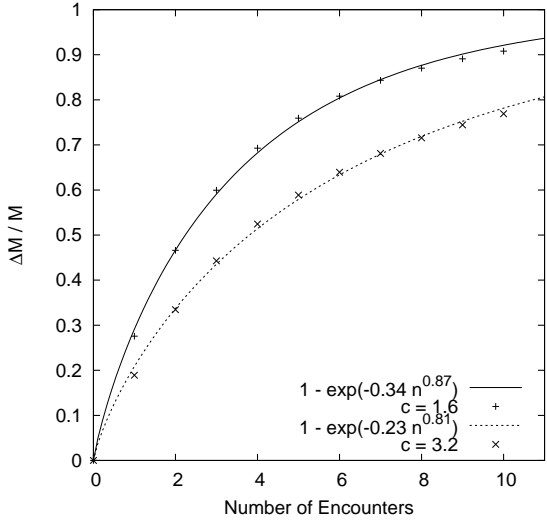}
\end{center}
\caption{Cumulative mass loss of different micro-haloes after a number of
identical encounters with $b = 0.02$ pc.}
  \label{fig:dm_mult}
\end{figure}

The density profile of the micro-halo after $n$ encounters is shown in
Fig. \ref{fig:den_mult}.
Every encounter heats the system and strips some mass, but each successive
encounter is less and less effective. Fig. \ref{fig:dm_mult}
shows the cumulative mass loss after each such encounter. It can be described
by the following function:
\begin{equation}
  \frac{\Delta M}{M}(n) = 1 - \exp \left (A n^B \right),
\label{eq:n_in}
\end{equation}
with A = -0.34 and B = 0.87 for c = 1.6 and A = -0.23 and B = 0.81 for c = 3.2,
which are also shown in Fig. \ref{fig:dm_mult}.

\begin{figure}
\begin{center}
\epsfxsize=8.4cm
\epsffile{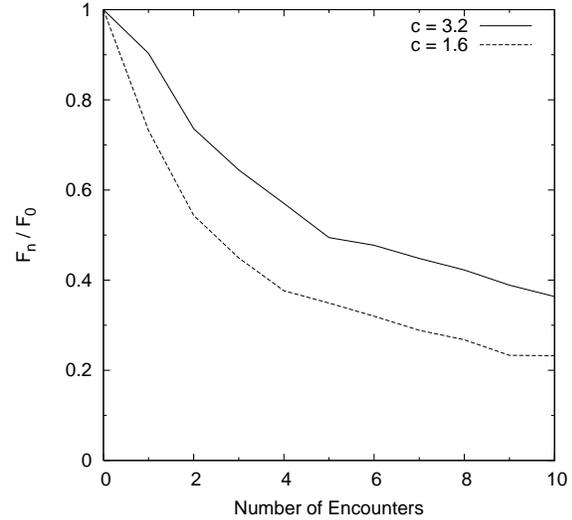}
\end{center}
\caption{The relative flux of annihilation products from different micro-haloes
after a given number of encounters. The typical mass loss from a halo 
would lead to a decrease in flux of between a factor of two or three.}
  \label{fig:flux}
\end{figure}

We fitted each of the eleven density profiles, which are shown in Fig.
\ref{fig:den_mult}, as well as the corresponding eleven density profiles for
the halo with c = 3.2, which are not shown, with equation (\ref{eq:den}). To
make the fits we used the Levenberg-Marquardt method \citep{marquardt} in
log-log space keeping $\gamma$ fixed at 1.2. The variation of $\alpha, \beta,
r_{\rm s}$ and $\rho_0$ from encounter to encounter is as follows: $\alpha$
increases from 1.0 to 1.5, $\beta$ increases monotonically from 3.0 to 7.0,
r$_{\rm s}$ stays constant around 7.0 milli-pc for the c = 1.6 halo and around
4.0 milli-pc for the c = 3.2 halo and finally $\rho_0$ oscillates around 20
GeV/cm$^3$ for c = 1.6 and decreases from 100 down to 50 GeV/cm$^3$ in case of
the other halo. The resulting profiles could now be used to calculate the net
flux coming from neutralino annihilation \citep[e.g,][]{savvas} via:
\begin{equation}
F = k \int^{\infty}_{r_{min}}{4 \pi r^2 \rho(r)^2 {\rm d} r}
\label{eq:annihil}
\end{equation}
We have summed up the dependence of the flux on neutralino mass and interaction
cross section in the constant $k$. The lower bound $r_{\rm min}$ is defined as
the central region of the micro halo, in which the neutralinos already
annihilated each other. The required number density for this to happen can be
estimated with the help of
\begin{equation}
t_{\rm h} = {1 \over n \sigma v}
\label{eq:lowbound}
\end{equation}
where $t_{\rm h} \approx 13$ Gyrs is the Hubble time, $\sigma v \approx
10^{-30}$ cm$^3$s$^{-1}$ is a typical cross section and $n$ is the number
density of neutralinos. For more details see \cite{calcaneo}. The minimum
radius can now be computed from comparing this minimum number density with the
density profile in Fig. \ref{fig:den_mult}. Assuming a neutralino mass of 100
GeV and deploying the above mentioned density profile, $r_{\rm min}$ comes out
to be $1.6 \times 10^{-14}$ pc. Fig. \ref{fig:flux}
shows then the resulting annihilation flux.

In order to investigate the sensitivity of tidal heating to the structure of
the micro-halo, we have run additional simulations with different initial
profiles.  We use the cuspy profile given by equation (\ref{eq:den}) and vary
the slope $\gamma = 0$, 0.5, 1, and 1.5, in addition to our fiducial value,
$\gamma=1.2$.  This suite of simulations is carried out using a fixed impact
parameter, $b=0.02$ pc, and other parameters as in the fiducial run. Our
calculations are similar to earlier studies of impulsive heating e.g.
\citet{aguilar86}, who studied the structural change in systems with de
Vaulouleur density profiles.  

\begin{figure}
\begin{center}
\epsfysize=8.4cm
\epsffile{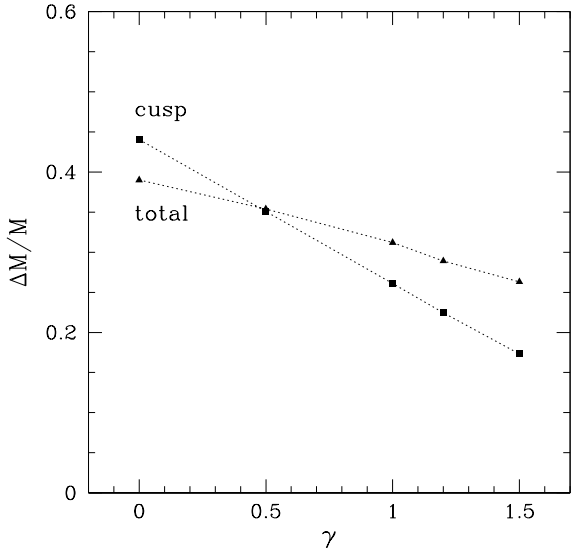}
\end{center}
\caption{Total mass loss from the micro-halo as a function of the
  inner density slope, $\gamma$, determined using the the position
  criterion, $r > R_{\rm max}$, after the potential relaxation ({\it
  triangles}), as well as the fraction of mass lost from inside the
  scale radius, $r < r_{\rm s}$ ({\it squares}).}
  \label{fig:dm_gam}
\end{figure}

Fig. \ref{fig:dm_gam}
shows that up to 30\% more mass is lost from the cored halo ($\gamma = 0$)
compared to the cuspy haloes ($\gamma > 1$). The effect is even stronger for
the fraction of mass removed from within the scale radius, $r_s$: 2.5 times
more material is lost from the cored halo.  The strongly-bound material within
the cusp is more stable against tidal disruption than that in cored profiles,
which have been typically considered in previous studies of tidal heating.

In Fig. \ref{fig:den_mult} we show the mass loss of the micro-halo for ten
successive encounters with exactly the same impact parameter. This is unlike 
the situation in our Galaxy where micro-haloes orbit the disc for 10 Gyrs near
the solar radius. It spends about 0.1 Gyr moving through the disc encountering
stars at a relative velocity of approximately 300\,km/s and a range of impact
parameters. In order to model this behaviour more precisely, we use a
Monte-Carlo method, which estimates the total amount of mass-loss it would
suffer. We draw encounter impact parameters from a random distribution and
calculate at each time the stripped mass. After each encounter, the density
profile of the micro-halo changes self-consistently according to the results
found earlier with the $N$-body simulations.

The mass loss due to the first encounter can easily be calculated using
equation \ref{eq:dm}. From the second encounter onwards, the halo density
profile changes and it is harder to strip during subsequent encounters as we
have seen in Fig. \ref{fig:dm_mult}. 

\begin{figure}
\begin{center}
\epsfysize=7.1cm
\epsffile{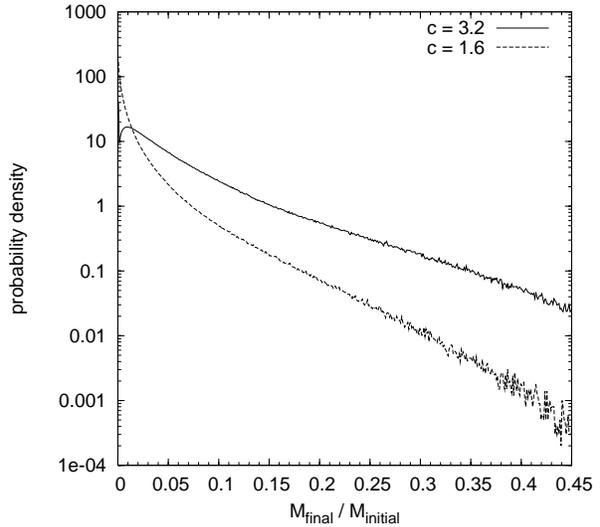}
\end{center}
\caption{Probability density distribution function of the final masses of
different haloes}
\label{fig:histo}
\end{figure}

The mass $M_{\rm a}$, the halo has after the $a$th encounter, determines the
reduction of the mass, which is stripped in the $a + 1$st encounter. This is
unfortunately not a variable of equation (\ref{eq:n_in}). This is a function of
$n$ only. So we calculate at the beginning of each encounter a ``virtual
encounter number'' $n$, which is basically the inverse of equation
(\ref{eq:n_in}):
\begin{equation}
n = \sqrt[B]{\log_{\rm e} M_{\rm a} \over A}
\end{equation}
For a given $M_{\rm a}$ it computes the corresponding number of ``standard
encounters'' with the impact parameter b\,=\,0.02\,pc from figure
\ref{fig:dm_mult}. Now we can calculate the mass $M_{\rm a + 1}$ after the $a +
1$st encounter. This must be a deviation from equation (\ref{eq:dm}) with some
weighting function $w(n,M_{\rm a})$:
\begin{equation}
M_{\rm a + 1} = M_{\rm a}\left[1 - {w(n,M_{\rm a})\over 1 + 2.1 \left(\Delta E
\over E_{\rm b} \right)^{-1}}\right],
\label{eq:mrma}
\end{equation}
This weighting function is the fraction of mass, which is stripped in the $n +
1$st standard encounter divided by the fraction of mass stripped in the first
standard encounter:
\begin{equation}
w(n,M_{\rm a}) = {{\Delta M \over M}(n + 1)-{\Delta M\over M}(n)\over{\Delta
M \over M}(1)\left(1 - {\Delta M \over M}(n)\right)}
\end{equation}
The ${\Delta M \over M}(n)$ can be calculated according to equation
(\ref{eq:n_in}). This weighting function reproduces the results from Fig.
\ref{fig:dm_mult}. Keeping in mind, that $\left[1-{\Delta M \over
M}(n)\right] = M_{\rm a}$, equation (\ref{eq:mrma}) reduces to:
\begin{equation}
M_{\rm a + 1} = M_{\rm a} - {M_{\rm a} - \exp \left[A(n+1)^B\right]
\over {\Delta M \over M}(1) \left[1 + 2.1 \left(\Delta E \over E_{\rm b}
\right)^{-1}\right]}
\end{equation}
We use this equation recursively for each encounter and then repeat the
calculation to obtain the probability distribution of final masses in Fig.
\ref{fig:histo}.
The central density (at our softening length) of a perturbed 
halo decreases only by a factor of about 5,
whilst the total mass decreases by an average of 
90 \% (see Fig. \ref{fig:den_mult}).

\section{Conclusions}

We have studied the disruption of dark matter micro-haloes by stars and other
substructures using both analytical impulse approximation and self-consistent
$N$-body simulations. The analytic calculations presented here are quite
similar to those of \citet{green_goodwin06} and we come to similar conclusions.
Our calculations differed in that we used more realistic cuspy n-body models
and we studied how the internal structure of these systems evolve due to
perturbations. See also \citet{angus} for an independent and complementary
study. Earlier studies  e.g. \citet{aguilar86}, also studied cuspy systems and
found similar robustness to tidal heating in the central regions as we find.
However our resolution allows us to study the response of CDM haloes deep
within their central regions.

\begin{itemize}

\item The impulse approximation predicts that those micro-haloes in the solar
vicinity which formed after z = 130 (about 80\% of the local micro-halo
population) should lose most of their mass due to 
close encounters with disc stars. 

\item Numerical simulations of individual encounters demonstrate that the usual
condition of disruptive heating used in analytical studies, $\Delta E = E_{\rm
b}$, does not lead to complete dissolution of haloes with cuspy density
profiles. For the inner logarithmic slope $\gamma = 1.2$, on average only 30\%
of the mass is lost from the system for this energy change. The relation
between the fractional mass loss and the energy input in the tidal
approximation is given by equation (\ref{eq:dm}): $\Delta M/M \approx (1 + 2
(\Delta E/E_{\rm b})^{-1})^{-1}$.

\item The change of particle energies, after the system settles into a new
virial equilibrium following the tidal encounter, is described accurately by
the extension of the impulse approximation accounting for virial oscillations.
An apparent resistance to tidal heating of the material deep in the cusp is
due to the high binding energy inside the cusp.

\item Repeated tidal encounters lead to diminishing mass loss from the same
micro-halo.  After 10 identical encounters at impact parameter $b = 2 R_{\rm
vir}$, 10\% of the halo still remains self-bound even though $\Delta E =
5E_{\rm b}$.

\end{itemize}

Near the solar radius within the Galaxy most of the mass of the micro-haloes
is tidally removed. This material forms cold streams in phase space providing a
unique signal for direct detection experiments. The dense cuspy cores of these
haloes survive reasonably intact, although the mass loss leads to a reduction
in annihilation products of about a factor of only two to three. These cores
could be distinguished by their high proper motions on the sky of the order
arc-minutes per year.

\section*{Acknowledgements}
For all $N$-body simulations we used  \textsc{Pkdgrav2} \citep{stadel}, a
multi-stepping tree code developed by Joachim Stadel. All computations were
made on the zBox supercomputer (www.zBox1.org) at the University of Z\"urich.
OYG acknowledges support from NASA ATP grant NNG04GK68G.

\label{lastpage}

\begin{thebibliography}{}

\bibitem[\protect\citeauthoryear{Aguilar \& White}{1985}]{aguilar85}
Aguilar L. A, White S. D. M, 1985, ApJ, 295, 374

\bibitem[\protect\citeauthoryear{Aguilar \& White}{1986}]{aguilar86}
Aguilar L. A, White S. D. M, 1986, ApJ, 307, 97

\bibitem[\protect\citeauthoryear{Angus \& Zhao}{2006}]{angus}
Angus G. W, Zhao H. S, 2006, astro-ph/0608580

\bibitem[\protect\citeauthoryear{Bahcall}{1984}]{bahcall84}
Bahcall J. N, 1984, ApJ, 276, 169

\bibitem[\protect\citeauthoryear{Berezinsky, Dokuchaev \& Eroshenko}{2006}]
{berezinsky}
Berezinsky V, Dokuchaev V, Eroshenko Y, 2006, Phys. Rev. D, 73, 063504

\bibitem[\protect\citeauthoryear{Binney \& Merrifield}{1998}]{BM98}
Binney J, Merrifield M, 1998, Galactic Astronomy (Princeton, NJ, 
  Princeton University Press), p. 130

\bibitem[\protect\citeauthoryear{Binney \& Tremaine}{1987}]{BT87}
Binney J, Tremaine S, 1987, Galactic Dynamics (Princeton, NJ, 
  Princeton University Press)

\bibitem[\protect\citeauthoryear{Boily et al.}{2004}]{boily}
Boily C. M, Nakasato N, Spurzem R, Tsuchiya T, 2004, ApJ, 614, 26 

\bibitem[\protect\citeauthoryear{Calc{\' a}neo-Rold{\' a}n \& Moore}{1999}]
{calcaneo}
Calc{\' a}neo-Rold{\' a}n C, Moore B, 2000, Phys. Rev. D, 62, 123005

\bibitem[\protect\citeauthoryear{Carr \& Sakellariadou}{1999}]
{carr_sakellariadou99}
Carr B. J, Sakellariadou M, 1999, ApJ, 516, 195

\bibitem[\protect\citeauthoryear{Diemand, Madau \& Moore}{2005}]{diemandmm} 
Diemand J, Madau P, Moore B, 2005, MNRAS, 364, 367

\bibitem[\protect\citeauthoryear{Diemand, Moore \& Stadel}{2004}]{diemand04} 
Diemand J, Moore B, Stadel J, 2004, MNRAS, 352, 535

\bibitem[\protect\citeauthoryear{Diemand, Moore \& Stadel}{2005}]{diemand05} 
Diemand J, Moore B, Stadel J, 2005, Nature, 433, 389

\bibitem[\protect\citeauthoryear{Diemand, Kuhlen \& Madau}{2006}]{diemand2006} 
Diemand J, Kuhlen M, Madau P, 2006, astro-ph/0603250

\bibitem[\protect\citeauthoryear{Flynn, Gould \& Bahcall}{1996}]{flynn_etal96}
Flynn C, Gould A, Bahcall J. N, 1996, ApJ, 466, 55 

\bibitem[\protect\citeauthoryear{Freudenreich}{1998}]{freudenreich98}
Freudenreich H. T, 1998, ApJ, 492, 495 

\bibitem[\protect\citeauthoryear{Gnedin \& Ostriker}{1999}]{gnedin_ostriker99}
Gnedin O. Y, Ostriker J. P, 1999, ApJ, 513, 626

\bibitem[\protect\citeauthoryear{Gould, Flynn \& Bahcall}{1998}]{gould_etal98}
Gould A, Flynn C, Bahcall J. N, 1998, ApJ, 503, 798 

\bibitem[\protect\citeauthoryear{Green \& Goodwin}{2006}]{green_goodwin06}
Green A. M, Goodwin S. P, 2006, astro-ph/0604142

\bibitem[\protect\citeauthoryear{Hernquist}{1990}]{abc}
Hernquist L, 1990, ApJ, 356, 359

\bibitem[\protect\citeauthoryear{Hofmann, Schwarz \& St\"ocker}{2001}]
{hofmann_etal01}
Hofmann S, Schwarz D. J, St\"ocker H, 2001, Phys. Rev. D, 64, 083507

\bibitem[\protect\citeauthoryear{Ivezi{\' c} et al.}{2000}]
{ivezic_etal00}
Ivezi{\' c} {\v Z}, Goldston J, Finlator K, Knapp G. R. et al, 2000, AJ, 120,
963 

\bibitem[\protect\citeauthoryear{Kazantzidis, Magorrian \& Moore}{2004}]
{kazantzidis_etal04}
Kazantzidis S, Magorrian J, Moore B, 2004, ApJ, 601, 37

\bibitem[\protect\citeauthoryear{Kazantzidis et al.}{2004}] {stelios}
Kazantzidis S, Mayer L, Mastropietro C, Diemand J, Stadel J, Moore B, 2004,
ApJ, 608, 663

\bibitem[\protect\citeauthoryear{Klypin, Zhao \& Somerville}{2002}]
{klypin_etal02}
Klypin A, Zhao H. S, Somerville R. S, 2002, ApJ, 573, 597

\bibitem[\protect\citeauthoryear{Koushiappas}{2006}]{savvas}
Koushiappas S. M, 2006, astro-ph/0606208

\bibitem[\protect\citeauthoryear{Kuijken \& Gilmore}{1989}]{kuijken_gilmore89}
Kuijken K, Gilmore G, 1989, MNRAS, 239, 605 

\bibitem[\protect\citeauthoryear{Marquardt}{1963}]{marquardt}
Marquardt D. W, 1963, Journal of the Society for Industrial and Applied
Mathematics, 11, 431

\bibitem[\protect\citeauthoryear{Moore}{1993}]{moore93}
Moore B, 1993, ApJ, 413, L93 

\bibitem[\protect\citeauthoryear{Moore et al.}{1999}]{moore_etal99}
Moore B, Quinn T, Governato F, Stadel J, Lake G, 1999, MNRAS, 310, 1147

\bibitem[\protect\citeauthoryear{Moore et al.}{2005}]{moore_etal05}
Moore B, Diemand J, Stadel J, Quinn T, 2005, astro-ph/0502213

\bibitem[\protect\citeauthoryear{Picaud \& Robin}{2004}]{picaud_robin04}
Picaud S, Robin A. C, 2004, A\&A, 428, 891 

\bibitem[\protect\citeauthoryear{Plummer }{1915}]{plummer}
Plummer H. C, 1915, MNRAS, 76, 107 

\bibitem[\protect\citeauthoryear{Spagna et al.}{2004}]{spagna_etal04}
Spagna A, Carollo D, Lattanzi, M. G, Bucciarelli B, 2004, A\&A, 428, 451

\bibitem[\protect\citeauthoryear{Spitzer}{1987}]{spitzer87}
Spitzer L, 1987, Dynamical Evolution of Globular Clusters (Princeton, NJ,
  Princeton University Press)

\bibitem[\protect\citeauthoryear{Stadel}{2001}]{stadel}
Stadel J, 2001, PhD thesis, Univ. Washington

\bibitem[\protect\citeauthoryear{Zhao et al.}{2005a}]{zhao_etal05a}
Zhao H. S, Hooper D, Angus G. W, Taylor J, Silk J, 2005a, astro-ph/0508215

\bibitem[\protect\citeauthoryear{Zhao et al.}{2005b}]{zhao_etal05b}
Zhao H. S, Taylor J, Silk J, Hooper D, 2005b, astro-ph/0502049

\end{thebibliography}
\end{document}